\documentclass[11pt]{article}

\usepackage{amsmath,amsfonts,amssymb}

\usepackage[numbers,sort&compress]{natbib}
\usepackage{graphicx}

\usepackage{hyperref}
\hypersetup{
   colorlinks   =  true,
    linkcolor    = blue,
    citecolor    = red,
     urlcolor	=magenta,     
}

\newcommand\be{\begin{equation}}
\newcommand\ee{\end{equation}}
\newcommand\bea{\begin{eqnarray}}
\newcommand\eea{\end{eqnarray}}

\DeclareMathOperator\spn{span}

 \newcommand\nullp{{\rm N}}
 
 \newcommand\st{{\rm st} }
 \newcommand\til{{\rm tl}} 
 \newcommand\sil{{\rm sl}} 
 \newcommand\lil{{\rm ll} }

 \newcommand\timek{{\rm time}}
\newcommand\spa{{\rm space}}
\newcommand\nullplane{{\rm light}}

\def\>#1{{\bf #1}} 
 \def\mf {\mathfrak}

\begin{document}

\ 

 \begin{center}
\noindent
 {\LARGE \bf  
{The noncommutative space of light-like worldlines
}} 
 \end{center}

\medskip

\begin{center}

{\sc Angel Ballesteros$^{1}$, Ivan Gutierrez-Sagredo$^{2}$ and Francisco J.~Herranz$^{1}$}

\medskip
{$^1$Departamento de F\'isica, Universidad de Burgos, 
09001 Burgos, Spain}

{$^2$Departamento de Matem\'aticas y Computaci\'on, Universidad de Burgos, 
09001 Burgos, Spain}
\medskip
 
e-mail: {\href{mailto:angelb@ubu.es}{angelb@ubu.es}, \href{mailto:igsagredo@ubu.es}{igsagredo@ubu.es}, \href{mailto:fjherranz@ubu.es}{fjherranz@ubu.es}}

\end{center}

\medskip

\begin{abstract}

The noncommutative space of light-like worldlines that is covariant under the light-like (or null-plane) $\kappa$-deformation of the (3+1) Poincar\'e group  is fully constructed as the quantization of the corresponding Poisson homogeneous space of null geodesics. This new noncommutative space of geodesics is five-dimensional, and turns out to be defined as a quadratic algebra that can be mapped to a non-central extension of the direct sum of two Heisenberg--Weyl algebras whose noncommutative parameter is just the Planck scale parameter $\kappa^{-1}$. Moreover, it is shown that the usual time-like $\kappa$-deformation of the Poincar\'e group does not allow the construction of the Poisson homogeneous space of light-like worldlines. Therefore, the most natural choice in order to model the propagation of massless particles on a quantum Minkowski spacetime seems to be provided by the light-like $\kappa$-deformation.

  \end{abstract}
\medskip
\medskip


\bigskip



\tableofcontents

\newpage


\section{Introduction}

Noncommutative spacetimes have been proposed in different approaches to Quantum Gravity as useful algebraic tools in order to describe the minimal length scenarios and the spacetime fuzziness that is believed to arise as a consequence of quantum properties of spacetime at the Planck scale (see, for instance,~\cite{Snyder1947, Mead:1964zz, Amati:1988tn, Ahluwalia:1993dd, Maggiore1993gup, DFR1994, Garay1995, MW1998,Hossenfelder2013minlength} and references therein). In this context, the covariance properties of such ``quantum'' spacetimes becomes a relevant issue, which can be solved in the case of the noncommutative analogues of spacetimes that in the classical setting can be obtained as homogeneous spaces of kinematical groups, as it is the case of Minkowski and (Anti-)de Sitter spacetimes. In fact, these quantum homogeneous spacetimes can be defined as covariant objects under a quantum kinematical group of symmetries, and they can be explicitly obtained as quantizations of classical Poisson homogeneous spaces. Moreover, the latter are covariant under a Poisson--Lie group, which is just the semiclassical counterpart of the corresponding quantum group~\cite{Drinfeld1987icm,ChariPressley1994}.

In this setting, a noncommutative (Minkowski or (A)dS) spacetime is mathematically defined as a comodule algebra under the action of the corresponding quantum group of transformations. In this way, the latter  provides the deformed analogue of the classical relativistic symmetries for the quantum spacetime, and allows the connection with the Deformed Special Relativity (DSR) approach to quantum gravity phenomenology in which the quantum deformation parameter is assumed to be related to the Planck scale~\cite{Amelino-Camelia2001testable,Kowalski-Glikman2001,Amelino-Camelia2002planckian,MS2002,LN2003versus,Freidel2004,Amelino-Camelia2010symmetry}.  So far, the most studied noncommutative spacetime model arising in this framework is the so-called $\kappa$-Minkowski spacetime (see~\cite{Maslanka1993,BRH2003,Lukierski2006,Daszkiewicz2008,BorowiecPachol2009jordanian,BP2014extendedkappa,BRH2017,PoinDD,BGH2019kappaAdS3+1,kappaN,GH2021symmetry,BGH2021lorentz} and references therein), which is covariant under the (time-like) quantum deformation known as the $\kappa$-Poincar\'e algebra~\cite{LRNT1991,GKMMK1992,LNR1992fieldtheory,Maslanka1993,MR1994,Zakrzewski1994poincare}. Also, its non-vanishing cosmological constant counterpart, the $\kappa$-(A)dS noncommutative spacetime, has also been  recently constructed~\cite{BGH2019kappaAdS3+1} from the corresponding $\kappa$-(A)dS quantum group introduced in~\cite{BHMN2017kappa3+1}.

However, there exists another class of very relevant homogeneous spaces for kinematical groups whose noncommutative analogues had not been considered until recently despite their physical significance: the homogeneous spaces of geodesics on the above-mentioned Lorentzian spacetimes with constant curvature. In fact, the construction of ``quantum geodesics" on a given noncommutative spacetime is the essential problem underlying the definition of ``quantum observers". The explicit construction of the noncommutative space of time-like worldlines associated to the $\kappa$-Minkowski spacetime has been recently given in~\cite{BRH2017, BGH2019worldlinesplb}, and the fuzziness of events has also  been  analysed in this quantum worldline setting~\cite{BGGM2021fuzzy}.

The aim of this paper is the explicit construction of the noncommutative space of light-like geodesics associated to the $\kappa$-deformation. Indeed, the study of the propagation of photons on a quantum spacetime is one of the main theoretical problems in quantum gravity phenomenology (see~\cite{Amelino-Camelia:2008aez,Addazi:2021xuf}) and we hope that this new model could shed some light in this context. Moreover, we will show that the construction of this new noncommutative space is possible only if the light-like (or null-plane) $\kappa$-deformation of the Poincar\'e algebra~\cite{nullplane95,Rnullplane97,bicrossnullplane} is considered. As we will show in detail, this comes from the fact that the usual time-like $\kappa$-Poincar\'e deformation is not mathematically compatible with the construction of a Poisson homogenous space of light-like geodesics through a canonical projection from the corresponding Poisson--Lie group. This result seems to indicate that the time-like and light-like $\kappa$-deformations could be considered separately in order to model the propagation of, respectively, time-like and light-like quantum particles.

The structure of the paper is as follows. In the next section we provide a general discussion about the conditions that a given classical $r$-matrix (and, therefore, its associated quantum deformation) has to present in order to allow a canonical construction of the Poisson homogeneous space whose quantization will provide the noncommutative space under consideration. Section~\ref{s3} provides the basics of the Poincar\'e group and its homogeneous spaces that will be needed in the rest of the paper, including the definition and properties of the so-called ``null-plane" basis, which will be essential in the rest of the paper. In section~\ref{s4}, the specific case of the homogenous space $\mathcal{L}$ of Minkowskian light-like geodesics will be analysed in detail from the viewpoint of the coisotropy condition described in section~\ref{s2}, thus arriving at the conclusion that the light-like $\kappa$-deformation is the only one that can be considered for the construction of the Poisson homogeneous space of light-like worldlines. The detailed construction of this space $\mathcal{L}$ is provided in section~\ref{s5}, where its quantization is performed, thus giving rise to the noncommutative space of light-like worldlines, whose algebraic properties are analysed in detail. A final section containing several remarks and open problems close the paper.


\section{From quantum deformations to noncommutative spaces}
\label{s2}

In general, the construction of a noncommutative analogue (endowed with quantum group invariance) of a given classical homogeneous space $G/H$ can be achieved through the following systematic and constructive procedure (see~\cite{Drinfeld1987icm,ChariPressley1994} and references therein):
\begin{itemize}

\item Select a given (coboundary) Poisson--Lie structure $\Pi$ on the Lie group $G$ with Lie algebra $ \mathfrak g$, which will be asociated to a unique $r$-matrix on $\mathfrak g\wedge \mathfrak g$. The Poisson   bivector $\Pi$ is explicitly given by means of the Sklyanin  bracket
\begin{align}
\begin{split}
\label{bb}
&\{f,g\}=r^{ij}\left( X^L_i f \, X^L_j g - X^R_i f \, X^R_j g \right),\qquad f,g \in \mathcal C (G),
\end{split}
\end{align} 
such that    $X^L_i$ and $ X^R_j$ are   left- and right-invariant vector fields on $G$. The quantization (as a Hopf algebra~\cite{Drinfeld1987icm,ChariPressley1994,majid,Abe}) of   
the Poisson--Lie group $(G,\Pi)$ is the corresponding quantum group.

\item Coboundary Lie bialgebras $(\mf g, \delta)$ are the tangent counterparts of coboundary Poisson--Lie groups $(G,\Pi)$, where the Lie bialgebra cocommutator map  $\delta: \mathfrak g\rightarrow \mathfrak g\wedge \mathfrak g$ is straightforwardly obtained from the $r$-matrix in the form
\be
\delta(X) = [X \otimes 1 + 1 \otimes X,r], \quad \forall X \in \mathfrak{g}.
\label{ba}
\ee

\item A Poisson homogeneous space $(G/H,\pi)$ for a  Poisson--Lie group $(G,\Pi)$  is the classical homogeneous space $G/H$ endowed with a Poisson structure $\pi$ which is covariant under the action of the Poisson--Lie group $(G,\Pi)$. The essential problem now is the explicit construction of the Poisson structure $\pi$, whose quantization will give rise to the noncommutative space with quantum group covariance we are looking for. In this respect, a strong mathematical constraint arises: such an explicit construction of $\pi$ is guaranteed only if the so-called coisotropy condition~\cite{Lu1990thesis,Ciccoli2006,BMN2017homogeneous}   for the cocommutator $\delta$ with respect to the isotropy subalgebra $\mathfrak h$ of $H$ holds, namely
 \be
\delta(\mathfrak h) \subset \mathfrak h \wedge \mathfrak g.
\label{coisotropy}
\ee
This condition can be interpreted as a strong compatibility constraint between the specific isotropy subgroup $H$ (and, therefore, between the chosen homogeneous space) and the specific Poisson--Lie structure for $G$ given by $\Pi$, which is in one to one correspondence with the cocommutator $\delta$. 
In the case that~\eqref{coisotropy} holds, we have a so-called coisotropic Poisson homogeneous space and the Poisson structure $\pi$ is straightforwardly obtained by canonical projection from the Poisson--Lie structure $\Pi$ on $G$, which can always be  explicitly constructed  in terms of the Sklyanin bracket~\eqref{bb}. Moreover, a  particular case of coisotropy condition~\eqref{coisotropy} is the so-called Poisson-subgroup condition, that holds when the Lie subalgebra $\mathfrak h $ is also a  sub-Lie bialgebra:  
\be
\delta\left(\mathfrak{h}\right) \subset \mathfrak{h} \wedge \mathfrak{h} ,
\label{coisotropy2}
\ee
which implies that $(H,\Pi |_H)$  is a  Poisson--Lie subgroup of  $(G,\Pi)$, \emph{i.e.}~$H$ is a Poisson submanifold of $G$.

\item Noncommutative spaces (which can be thought of as the generating objects of quantum homogeneous spaces~\cite{Dijkhuizen1994}) can then be obtained as the  quantization  (as a comodule algebra under the quantum group co-action)  of  coisotropic Poisson homogeneous spaces. In this quantum setting, the  coisotropy condition (\ref{coisotropy}) ensures that the commutation relations that define the noncommutative space  at the first-order in all the quantum coordinates generate a Lie subalgebra which is just the   annihilator $\mathfrak{h}^\perp$ of $\mathfrak{h}$ on the dual Lie algebra $\mathfrak g^*$ (see~\cite{Ciccoli2006,BGM2019coreductive,GH2021symmetry} for details).

 \end{itemize}   
 
As we will show in section~\ref{s4},  the most common  $\kappa$-Poincar\'e deformation (namely, the time-like $\kappa$-Poincar\'e algebra~\cite{LRNT1991,GKMMK1992,LNR1992fieldtheory,Maslanka1993,MR1994,Zakrzewski1994poincare} ) does not fulfil the coisotropy condition for the homogeneous space of null-geodesics on Minkowski spacetime, and therefore the previous construction will be precluded. In contradistinction, we will show that the coisotropy condition  is fulfilled by the light-like $\kappa$-Poincar\'e algebra~\cite{nullplane95,Rnullplane97,bicrossnullplane,BP2014extendedkappa} (which is also known in the literature as the ``null-plane" quantum Poincar\'e algebra), which therefore becomes the natural candidate as the quantum symmetry of the non-commutative space of null geodesics, which will be explicitly constructed in section~\ref{s5}.


\section{The (3+1) Poincar\'e group and its homogeneous spaces}
\label{s3}

Let us consider the Poincar\'e Lie algebra $\mathfrak{g}=  \mathfrak{iso}(3,1)  \equiv \mathfrak{so}(3,1) \ltimes \mathbb{R}^4$, which generates the (3+1)D Poincar\'e group $G =  {\rm  ISO}(3,1)$. In the usual kinematical basis $\{P_0,P_a, K_a, J_a\}$  $(a=1,2,3)$ of generators of time translation, space translations, boosts and rotations, respectively, the commutation rules for $\mathfrak{g}$ read
\be
\begin{array}{llll}
[J_a,J_b]=\epsilon_{abc}J_c ,& \quad [J_a,P_b]=\epsilon_{abc}P_c , &\quad
[J_a,K_b]=\epsilon_{abc}K_c ,  &\quad  [J_a,P_0]=0 , \\[2pt]
\displaystyle{
  [K_a,P_0]=P_a  } , &\quad\displaystyle{[K_a,P_b]=\delta_{ab} P_0} ,    &\quad\displaystyle{[K_a,K_b]=-\epsilon_{abc} J_c} , 
 &\quad 
[P_\mu,P_\nu]=0 ,
\end{array}
\label{aa}
\ee
where the speed of light is set $c=1$.
From now on sum over repeated indices will be assumed, unless otherwise stated. In the above kinematical basis, $a,b,c=1,2,3$, and $\mu,\nu=0,1,2,3$. We shall denote  3-vectors by $\mathbf{v}=(v^1,v^2,v^3)$ and 4-vectors by $ {v}=(v^0,\mathbf{v})= (v^0,v^1,v^2,v^3)$.

A faithful representation $\rho : \mathfrak g  \rightarrow \text{End}(\mathbb R ^5)$ for a generic  element $X\in \mathfrak g $ is given by 
\begin{equation}\rho(X)=   x^\mu \rho(P_\mu)  +  \xi^a \rho(K_a) +  \theta^a \rho(J_a) =\left(\begin{array}{ccccc}
0&0&0&0&0\cr 
x^0 &0&\xi^1&\xi^2&\xi^3\cr 
x^1 &\xi^1&0&-\theta^3&\theta^2\cr 
x^2 &\xi^2&\theta^3&0&-\theta^1\cr 
x^3 &\xi^3&-\theta^2&\theta^1&0
\end{array}\right) \, ,
\label{ab}
\end{equation}
and the corresponding exponential map provides a 5D representation the Poincar\'e group $G$.
 
Let us also introduce the so called null-plane basis for the Poincar\'e algebra~\cite{Leutwyler78}, which was   the one used in the construction of the  ``null-plane" quantum Poincar\'e algebra ~\cite{nullplane95,Rnullplane97,bicrossnullplane}, since this basis will be essential for our construction. We consider the null-plane $\nullp_n^\tau$ orthogonal to the light-like vector $n=(\frac 12,0,0,\frac 12)$ in the classical Minkowski spacetime with   Cartesian coordinates $x$  and we define
\be
x^+=x^0+x^3,\qquad x^-=\frac 12(x^0-x^3) =\tau.
\label{ac}
\ee
A point $x\in \nullp_n^\tau$  is labelled by the coordinates $(x^+,x^1,x^2)$, while $x^-=\tau$   plays the role of a ``time" or evolution parameter, and the chosen null-plane is associated with the boost generator $K_3$. It can be checked that  under the action of the boost transformation generated by the $K_3$ uniparametric subgroup,   the initial null-plane $\nullp_n^0$ $(x^-=0)$ is invariant and  the transverse coordinates $(x^1,x^2)$ remain unchanged, while $\exp(\xi^3 \rho(K_3))$ maps $x^+$ into ${\rm e}^{\xi^3} x^+$. The generators of the Poincar\'e algebra (\ref{aa}) can be sorted into three different classes  according to the adjoint action of $K_3$ onto them, namely
\be
[K_3,X]=\gamma X,\qquad X\in \mf g ,
\label{goodness}
\ee
where the parameter $\gamma$ is called  the ``goodness" of the generator $X$~\cite{Leutwyler78}. Such property allows us to introduce a null-plane  Poincar\'e algebra basis associated with the coordinates (\ref{ac}), which is consistent with all the results already presented in~\cite{nullplane95,Rnullplane97,bicrossnullplane}. Explicitly, the sign of the rotation generators has to be changed
\be
L_a=-J_a ,
\label{aad}
\ee
and the so-called ``null-plane basis"  $\{P_\pm,P_i, E_i, F_i,K_3,L_3\}$ $(i=1,2)$ is defined as
\be
\begin{array}{ll}
\mbox{$\gamma=+1$:}&\quad \displaystyle{ P_+=\frac 12 (P_0+P_3) ,\quad  E_1=\frac 12(K_1+L_2)  ,\quad E_2=\frac 12(K_2 - L_1). }\\[6pt]
\mbox{$\gamma=0$:}&\quad   K_3,\quad L_3,\quad P_1,\quad P_2 . \\[5pt]
\mbox{$\gamma=-1$:}&\quad  P_-= 
P_0-P_3,\quad F_1=  K_1
-L_2 ,\quad F_2= K_2 + L_1.
\end{array}
\label{ad}
\ee
From the commutation relations (\ref{aa}) we obtain that~\cite{nullplane95,bicrossnullplane}
\be
\begin{array}{llll}
 [L_3,E_i]=-\epsilon_{ij3}E_j,& \quad [L_3,F_i]=-\epsilon_{ij3}F_j , &\quad
[L_3,P_i]=-\epsilon_{ij3}P_j ,  &\quad   [L_3,P_\pm]=0 , \\[2pt]
    [K_3,E_i]=E_i  , &\quad [K_3,F_i]=-F_i,    &\quad [K_3,P_i]=0, 
 &\quad 
[K_3,P_\pm]=\pm P_\pm,\\[2pt]
 [E_i,P_j]=\delta_{ij}P_+, &\quad [E_i,P_+]=0,    &\quad  [E_i,P_-]=P_i, 
 &\quad 
[E_1,E_2]=0,\\[2pt]
 [F_i,P_j]=\delta_{ij}P_-, &\quad [F_i,P_+]= P_i,    &\quad   [F_i,P_-]= 0, 
 &\quad 
[F_1,F_2]=0,\\[2pt]
[K_3,L_3]=0, 
 &\quad 
[P_\alpha,P_\beta]=0,&    \multicolumn{2}{l} {\quad
 [E_i,F_j]=\delta_{ij}K_3 +\epsilon_{ij3}L_3,    }  
 \end{array}
\label{ae}
\ee
where the indices in the null-plane basis are $i,j=1,2$ and $\alpha, \beta=\pm,1,2$.

From the kinematical representation (\ref{ab}) we obtain the corresponding matrices for the null-plane basis generators (\ref{aad}) and (\ref{ad}), and the generic   element $X\in \mathfrak g $ reads~\cite{Rnullplane97}
\bea
&&  \rho(X)=   x^+ \rho(P_+) + x^- \rho(P_-) + x^i  \rho(P_i)+ e^i  \rho(E_i)+ f^i  \rho(F_i)+\xi^3  \rho(K_3)+\phi^3  \rho(L_3) \nonumber\\[4pt]
&&\qquad\  \, =\left(\begin{array}{ccccc}
0&0&0&0&0\\[2pt]
\frac 12 x^+ + x^- &0&\frac12 e^1 +f^1 &\frac12 e^2 +f^2&\xi^3\\[2pt]
x^1 &\frac12 e^1 +f^1&0&\phi^3&-\frac12 e^1 +f^1\\[2pt]
x^2 &\frac12 e^2 +f^2&-\phi^3&0&-\frac12 e^2 +f^2\\[2pt]
\frac 12 x^+ - x^-&\xi^3&\frac12 e^1 -f^1&\frac12 e^2 -f^2&0
\end{array}\right) \, .
\label{af}
\eea
 
From the representation (\ref{af}) it can be directly checked   that the seven generators with $\gamma=+1$ and $\gamma=0$ span the isotropy subgroup of the initial null-plane $\nullp_n^0$,  keeping $x^-=0$,   while the three remaining ones with 
$\gamma=-1$ act non-trivially on $\nullp_n^0$. In particular,  the transformations generated by $F_i$ rotate $\nullp_n^0$, and $\exp(\tau\rho( P_-))$
transforms   $\nullp_n^0$ into $\nullp_n^{\tau}$. Hence the latter generators, which span an abelian subgroup, determine the dynamical evolution of  $\nullp_n^0$ with time $x^-=\tau$.

It can also  be  shown that the  five generators $\{P_+,E_i,K_3,L_3\}$ span the isotropy subgroup of the light-like geodesic determined by $n=(\frac 12,0,0,\frac 12)$ which under
    the $5\times 5$ matrix representation (\ref{af}) is associated to the 5-vector
\be
\ell=\left(1,\tfrac 12,0,0,\tfrac 12 \right).
\label{ag}
\ee
This means that
 \begin{align}
\begin{split}
&\exp\left(  x^+ \rho(P_+) \right)\exp\left(  e^1 \rho(E_1) \right)\exp\left(  e^2  \rho(E_2) \right)\exp\left(\xi^3  \rho(K_3)  \right)\exp\left( \phi^3  \rho(L_3) \right)  \ell^T\\
&\qquad \qquad=
 \left(  1, \tfrac 12 {\rm e}^{\xi^3}\!+\tfrac 12 x^+,0,0, \tfrac 12 {\rm e}^{\xi^3}\!+\tfrac 12 x^+ \right)^T ,
 \end{split}
 \label{ah}
\end{align}
where $T$ denotes transpose. The transformations generated by the five remaining generators  $\{P_-,P_i,F_i\}$ can be used to transform $\ell$ (\ref{ag}) into any other light-like worldline, and they generate a subgroup which can be thought of as a central extension through $P_-$  of the 4D abelian group. Therefore, the space of light-like worldlines will have dimension 5.
 
Finally, we recall that, as a vector space, the Poincar\'e algebra $\mathfrak g$ can be decomposed in the following ways:
\be
\begin{array}{lll}
\mathfrak g= \mathfrak t_\st \oplus \mathfrak h_\st ,  \quad & \mathfrak{t}_\st = \spn \{P_0, \mathbf{P} \}    ,\  & \mathfrak h_\st = \spn\{\mathbf{K}, \mathbf{J} \}=\mathfrak{so}(3,1) ,\\[4pt]
\mathfrak g= \mathfrak t_\til \oplus \mathfrak h_\til ,  \quad & \mathfrak{t}_\til = \spn \{ \mathbf{P}, \mathbf{K} \}    ,\   & \mathfrak h_\til = \spn\{P_0, \mathbf{J} \}=\mathbb R\oplus \mathfrak{so}(3) ,\\[4pt]
\mathfrak g= \mathfrak t_\sil \oplus \mathfrak h_\sil ,  \quad & \mathfrak{t}_\sil = \spn \{P_0,P_i,  K_3,J_i\}    ,\   & \mathfrak h_\sil = \spn\{P_3,  K_i,J_3 \}=\mathbb R\oplus \mathfrak{so}(2,1) , \\[4pt]
\mathfrak g= \mathfrak t_\lil \oplus \mathfrak h_\lil ,  \quad & \mathfrak{t}_\lil = \spn \{P_-,P_i,F_i\}    ,\   & \mathfrak h_\lil = \spn\{P_+,E_i,K_3,L_3 \}  ,  
\end{array}
\label{ai} 
\ee
where here $i=1,2$ and  `$\st$', `$\til$', `$\sil$' and `$\lil$' means, in this order, spacetime, time-like, space-like and light-like. From these decompositions the corresponding homogeneous spaces of points and lines  can be constructed as the 
 left coset spaces  $G/H$ between the Poincar\'e group $G$ and the isotropy subgroup $H$ with Lie algebra $\mathfrak h$ (\ref{ai}), namely
 \begin{itemize}
  \item The (3+1)D Minkowski spacetime is obtained as the homogeneous space $\mathcal{M}  =G/H_\st$, where the isotropy subgroup $H_\st$ is just the Lorentz group.
 
 \item The 6D space of time-like lines is obtained as the homogeneous space $\mathcal{W}_\til  =G/H_\til $.
 
 \item The 6D space of space-like lines would be given by $\mathcal{W}_\sil  =G/H_\sil$. 

 \item And, finally, the 5D space of light-like lines can be constructed as $ \mathcal{L}  =G/H_\lil$.

 \end{itemize}   
 
The aim of this paper is the construction of a noncommutative version of $\mathcal{L}$ which is covariant under a quantum Poincar\'e group.


\section{$\kappa$-Poincar\'e deformations and light-like worldlines} 
\label{s4}

We recall that all   Lie bialgebra structures for the (3+1)D Poincar\'e     algebra $\mf g$   are coboundary ones~\cite{Zakrzewski1995,Zakrzewski1997,PW1996}. Therefore any quantum deformation is determined by an underlying classical $r$-matrix which  provides the cocommutator $\delta$ in the form~\eqref{ba}.

Among all the possible inequivalent $r$-matrices for the Poincar\'e algebra (see~\cite{Zakrzewski1997} for a classification), the most studied cases are the so-called $\kappa$-Poincar\'e $r$-matrices, which in covariant notation can be written as (see~\cite{Zakrzewski1997, BP2014extendedkappa})  
\be
r= a^{\mu}\,  M_{\mu\nu}\wedge P_\nu\, ,
\ee
where $a^{\mu}$ are the components of a Minkowskian four-vector $a$ and $M_{\mu\nu}$ denote the generators of the Lorentz group with the following identifications:
\be
M_{0\nu} \equiv K_\nu,\qquad
M_{12}= J_3,
\qquad
M_{23}= J_1,
\qquad
M_{31}= J_2.
\ee

In this way, three neat classes of $\kappa$-deformations arise:
\begin{itemize}

\item The time-like deformation: if we consider the time-like vector $a=(1/\kappa,0,0,0)$, we obtain
\be
r_\timek= \frac{1}{\kappa} (M_{0\nu}\wedge P_\nu) = \frac 1\kappa \left(K_1\wedge P_1 +K_2\wedge P_2 + K_3\wedge P_3 \right) .
\label{rtime}
\ee

\item The space-like deformation, generated by the space-like vector $a=(0,0,0,-1/\kappa)$:
\be
r_\spa= -\frac{1}{\kappa} (M_{3\nu}\wedge P_\nu) =  \frac 1\kappa\left(K_3\wedge P_0 +J_1\wedge P_2 -J_2\wedge P_1 \right) .
\label{rspace}
\ee

\item The light-like deformation: obtained from the light-like vector $a=(1/\kappa,0,0,-1/\kappa)$:
\bea
&& r_\nullplane= \frac{1}{\kappa} (M_{0\nu}\wedge P_\nu - M_{3\nu}\wedge P_\nu)\nonumber\\[2pt]
&&\qquad \  =  \frac 1\kappa\left(K_1\wedge P_1 +K_2\wedge P_2 + K_3\wedge P_3 + K_3\wedge P_0 +J_1\wedge P_2 -J_2\wedge P_1 \right).
\label{rlight}
\eea

\end{itemize}

 The first $r$-matrix, $r_\timek$, is just the  $r$-matrix underlying the well-known $\kappa$-Poincar\'e algebra~\cite{LRNT1991,GKMMK1992,LNR1992fieldtheory,Maslanka1993,MR1994,Zakrzewski1994poincare}, whose time-like nature is not frequently emphasized in the literature. In this case, the deformation parameter $\kappa$ has  dimensions of a ${\rm time}^{-1}$ (recall that $c=1$). 
The second one, $r_\spa$, determines the   $q$-Poincar\'e algebra obtained in~\cite{CK4d} (c.f.~Type 1.~(a) with  $q={\rm e}^{z}$ and $z=1/\kappa$),  having  $\kappa$   dimensions of a ${\rm length}^{-1}$. Both $r_\timek$ and $r_\spa$ provide quasitriangular (or standard) deformations of the Poincar\'e algebra since they are solutions of the modified classical Yang--Baxter equation with non-vanishing Schouten bracket. In fact,  both $r$-matrices/deformations can be obtained through coboundary Lie bialgebra contractions from  the Drinfel'd--Jimbo   bialgebra for $\mathfrak{so}(5)$ (see~\cite{GH2021symmetry} and references therein).

Finally, we stress that $r_\nullplane$ gives rise exactly to the null-plane quantum Poincar\'e algebra
 introduced in~\cite{nullplane95,Rnullplane97,bicrossnullplane} (where $z=1/\kappa$). This is a triangular (or nonstandard) deformation 
 with vanishing  Schouten bracket, which
indeed, verifies
  \be
  r_\nullplane=r_\timek+r_\spa .
  \label{bc}
  \ee
In  Table~\ref{table1} the three $r$-matrices are written in both the kinematical (\ref{aa}) and null-plane basis (\ref{ae}). As it can be easily appreciated, the latter is very well adapted to the light-like deformation, while the kinematical basis provides the simplest possible form for  both the time-like and space-like deformations.

Now let us face the explicit construction of the noncommutative analogue of the homogeneous space of light-like worldlines $ \mathcal{L}  =G/H_\lil$ where 
\be
\mathfrak h_\lil = \spn\{P_+,E_i,K_3,L_3 \}.
\label{hll}
\ee
As it was explained in section~\ref{s2}, firstly we have to obtain a coisotropic Poisson homogeneous space $(\mathcal{L} ,\pi)$ where $\pi$ is the canonical projection of a Poisson--Lie structure $\Pi$ on the Poincar\'e group generated by a given classical $r$-matrix, which has an associated cocommutator map $\delta$ determined by~\eqref{ba}. But this construction is only possible if the coisotropy condition with respect to the isotropy subalgebra $\mathfrak h_\lil $~\eqref{hll} of the space of light-like geodesics is fulfilled, namely:
 \be
\delta(\mathfrak h_\lil) \subset \mathfrak h_\lil \wedge \mathfrak g \, .
\label{coisotropyll}
\ee
A straightforward computation shows that the light-like $r$-matrix~\eqref{rlight} does fulfill this condition, while both the time-like~\eqref{rtime} and the space-like~\eqref{rspace} $r$-matrices do not. Explicitly, the cocommutator obtained from the light-like $r$-matrix~\eqref{rlight},  via the relation (\ref{ba}), reads
\bea
&&\delta(X)=0,\quad X\in\{P_+,E_i,L_3\}, \nonumber\\ 
&&\delta(Y)=\frac 2\kappa\, Y\wedge P_+,\quad Y\in\{P_-,P_i\},\nonumber\\
&&\delta(K_3)=\frac 2\kappa\,  (K_3\wedge P_++E_1\wedge P_1+E_2\wedge P_2),\label{ca}\\
&&\delta(F_1)=\frac 2\kappa\, (F_1\wedge P_++E_1\wedge P_- +L_3\wedge P_2),\nonumber\\
&&\delta(F_2)=\frac 2\kappa\,  (F_2\wedge P_++E_2\wedge P_- -L_3\wedge P_1) \, .
\nonumber
\eea
Therefore~\eqref{rlight} generates the unique $\kappa$-deformation that can provide a coisotropic Poisson homogeneous space of null geodesics, whose explicit construction will be presented in the following section, together with its quantization.

  
\begin{table}[t]
{\small
\caption{\small   The classical $r$-matrices  for the time-like, space-like and light-like   $\kappa$-deformations of the (3+1)D Poincar\'e algebra in both the kinematical (\ref{aa}) and null-plane basis (\ref{ae}).}
\label{table1}
  \begin{center}
\noindent 
\begin{tabular}{ l l}
\hline

\hline
\\[-6pt]
$\bullet$ $ r_\timek$ &\!\!\!\!\!$= \frac 1\kappa \left(K_1\wedge P_1 +K_2\wedge P_2 + K_3\wedge P_3 \right) $  \\[4pt]
&\!\!\!\!\!$=  \frac 1\kappa\left(K_3\wedge P_+ +E_1\wedge P_1+ E_2\wedge P_2 \right) -\frac {1}{2\kappa}\left(K_3\wedge P_- -F_1\wedge P_1-F_2\wedge P_2 \right) $  \\[6pt]
$\bullet$ $ r_\spa$ &\!\!\!\!\!$=  \frac 1\kappa\left(K_3\wedge P_0 +J_1\wedge P_2 -J_2\wedge P_1 \right)$  \\[4pt]
&\!\!\!\!\!$=  \frac 1\kappa\left(K_3\wedge P_+ +E_1\wedge P_1+ E_2\wedge P_2 \right) +\frac {1}{2\kappa}\left(K_3\wedge P_- -F_1\wedge P_1-F_2\wedge P_2 \right)$  \\[6pt]
$\bullet$ $ r_\nullplane$ &\!\!\!\!\!$=   \frac 1\kappa\left(K_3\wedge P_0 +K_1\wedge P_1 +K_2\wedge P_2 + K_3\wedge P_3 +J_1\wedge P_2 -J_2\wedge P_1 \right)  $  \\[4pt]
&\!\!\!\!\!$= \frac 2\kappa\left(K_3\wedge P_+ +E_1\wedge P_1+ E_2\wedge P_2 \right)$  \\[6pt]
\hline

\hline
\end{tabular}
 \end{center}
}
 \end{table} 
 


\section{Construction of the noncommutative space}
\label{s5}

In order to construct the noncommutative space of light-like worldlines we will follow a similar approach to the one presented in~\cite{BGH2019worldlinesplb} for the construction of its time-like analogue. We start by considering the following Poincar\'e group element obtained from the matrix realization (\ref{af}):
\be
G_\mathcal{L}= \exp\left( {y^- \rho(P_-)} \right)\exp\left({f^1 \rho(F_1)}  \right)\exp\left( {f^2 \rho(F_2)} \right)  \exp\left({y^1 \rho(P_1)}  \right)\exp\left( {y^2 \rho(P_2)}  \right)  H_\lil  ,
 \label{ce}
 \ee
where  $H_\lil$ is the isotropy subgroup or stabilizer of the light-like line $\ell$  (\ref{ag}), since the latter is taken as the origin of the 5D homogeneous space $\mathcal{L}  =G/H_\lil$ (see (\ref{ah})). Explicitly,
\be
 H_\lil= \exp\left({y^+ \rho(P_+)}  \right)\exp\left({e^1 \rho(E_1)}  \right)\exp\left( {e^2 \rho(E_2)} \right)  \exp\left({\xi^3 \rho(K_3)} \right)  \exp\left({\phi^3 \rho(L_3)} \right)   .
\label{cf}
\ee 
We recall that under this parametrization the generators $\{ P_-, F_i,P_i\}$ become the generators of translations on the 5D classical coset space $\mathcal{L} $, that is parametrized in terms of the $(y^-, f^i,y^i)$ coordinate functions. Note also that the order chosen
 for calculating the group element $G_\mathcal{L}$  (\ref{ce}) is consistent with the coset structure induced by the isotropy subgroup of the null-plane, which is generated by $\{ P_i, P_+, E_i, K_3, L_3  \}$.

From (\ref{ce}),  a long but straightforward computation leads to the left- and right-invariant vector fields on $G$, and afterwards the full Poisson--Lie group structure $\Pi$ on the Poincar\'e group corresponding to the light-like $\kappa$-deformation can be explicitly obtained by means of the Sklyanin  bracket (\ref{bb}) for the  $r$-matrix $r_\nullplane$~\eqref{rlight}. The non-vanishing Poisson brackets defining $\Pi$ are the following ones:
\be
\begin{array}{ll}
\displaystyle{ \big\{ y^-,  y^i \big\}=-\frac 2\kappa \,   f^i   y^- , }&\quad\displaystyle{ \big\{ y^-,  e^i \big\}= - \frac{1}{\kappa} \left( 2 f^i + e^i \bigl( ( f^1)^2+  ( f^2)^2\bigr) \right)  , }\\[8pt]
\displaystyle{ \big\{  y^-, \xi^3 \big\}=  - \frac{1}{\kappa} \bigl( ( f^1)^2+  ( f^2)^2\bigr) , }&\quad 
\displaystyle{ \big\{  y^+  , y^i\big\}= - \frac{2}{\kappa}\, y^i  , }\\[8pt]
\displaystyle{ \big\{ y^+ , e^i \big\}= - \frac{2}{\kappa}  \left( \exp{\xi^3} -1 \right) e^i , }&\quad \displaystyle{ \big\{ y^+,  f^i\big\}= -\frac{2}{\kappa}\, f^i, }\\[8pt]
\displaystyle{ \big\{ y^+ , \xi^3 \big\}= - \frac{2}{\kappa}  \left( \exp{\xi^3} -1 \right)  , } &\quad \displaystyle{ \big\{  y^i,  f^j \big\}= \frac 1\kappa \,\delta_{ij}   \bigl( ( f^1)^2+  ( f^2)^2\bigr) }, \\[8pt]
\displaystyle{ \big\{ y^1 , e^1 \big\}=  - \frac{2}{\kappa} \left( \exp{\xi^3} -1 - e^2 f^2 \right) , }&\quad \displaystyle{ \big\{ y^1 , e^2 \big\}= - \frac{2}{\kappa} \,e^1 f^2 , }\\[8pt]
\displaystyle{ \big\{ y^2 , e^2  \big\}=  - \frac{2}{\kappa} \left( \exp{\xi^3} -1 - e^1 f^1 \right)  , }&\quad \displaystyle{ \big\{ y^2 , e^1 \big\}= - \frac{2}{\kappa} \, e^2 f^1 , }\\[8pt]
\displaystyle{ \big\{ y^i , \phi^3  \big\}= \frac{2}{\kappa}\, \epsilon_{ij3} f^j   , }&\quad \displaystyle{ \big\{  y^1, y^2 \big\}= \frac 2\kappa\bigl(  f^1  y^2 -   f^2 y^1\bigr) , }\\[8pt]
\multicolumn{2}{l}{ \displaystyle{ \big\{ y^-,  y^+ \big\}=   \frac{1}{\kappa} \left( 2 y^- -  2 ( f^1 y^1 + f^2 y^2)  - y^+ \bigl( ( f^1)^2+  ( f^2)^2\bigr)  \right) . } }
 \end{array}
\label{chpois}
\ee

Now, the Poisson structure $\pi$ among the five  coordinates $(y^-,y^i,f^i )$ is obtained as the canonical projection of $\Pi$ onto this set of coordinates, which indeed generate a Poisson subalgebra. From~\eqref{chpois} it is immediate to check that the Poisson algebra generated by $(y^-,y^i,f^i )$ is quadratic homogeneous, and defines the Poisson homogeneous space $(\mathcal{L},\pi)$ associated to the light-like $\kappa$-deformation. Note that this result is fully consistent with~\eqref{ca} since the linearization of $\pi$ is zero, this means that at the r.h.s.~of the cocommutators~\eqref{ca} there should not exist any term $X\wedge Y$ with $X,Y\in  \{P_-,P_i,F_i\} $, and this is indeed the case. 

Moreover, since the Poisson algebra $\pi$ presents no ordering problems, it can  be directly quantized and the algebra $\mathcal{L}_\kappa$ so obtained defines the noncommutative space of light-like geodesics:
  \be
\begin{array}{ll}
\displaystyle{ \bigl[\hat y^-,\hat  y^i \bigr]=-\frac 2\kappa \, \hat  f^i \hat  y^- , }&\quad \displaystyle{ \bigl[\hat y^1,\hat y^2\bigr]= \frac 2\kappa\bigl( \hat f^1 \hat y^2 - \hat  f^2\hat y^1\bigr) , }\\[6pt] 
\displaystyle{ \bigl[\hat y^i, \hat f^j \bigr]= \frac 1\kappa \,\delta_{ij}   \bigl( (\hat f^1)^2+  (\hat f^2)^2\bigr) , }&\quad  \bigl[\hat y^-,\hat f^i \bigr]=  \bigl[\hat f^1,\hat f^2 \bigr]=0,\qquad i,j=1,2.
 \end{array}
\label{ch}
\ee
We remark that the three quantum coordinates $( \hat y^-,\hat f^i)$ can be thought to define  the quantum null-plane, which generates a 3-dimensional abelian subalgebra.

Let us  analyse the quantum space of light-like geodesics  $\mathcal{L}_\kappa$ (\ref{ch})   in more detail.
The three commutative quantum coordinates $( \hat y^-,\hat f^i)$ associated with the null-plane provide a (rational) Casimir operator $\hat C$ for the whole  $\mathcal{L}_\kappa$ algebra, which is given by
\be
\hat C = {\hat y^-}\left ( (\hat f^1)^2+  (\hat f^2)^2\right )^{-1} \, ,
\label{ci}
\ee
provided that $ \big( (\hat f^1)^2+  (\hat f^2)^2\big)^{-1}$ exists, and where we have made use of the expressions
$
\bigl[u,  v^{-1}\bigr] =-  v^{-1}  [u,  v  ]  \,v^{-1} ,
$ and the fact that $ \bigl[\hat y^-, (\hat f^1)^2+  (\hat f^2)^2 \bigr]=0$.

The four quantum coordinates $( \hat y^i,\hat f^i)$ generate a   subalgebra of $\mathcal{L}_\kappa$ which resembles a pair of two Heisenberg--Weyl algebras, and the commutator  $[\hat y^1,\hat y^2]$ in (\ref{ch}) gives rise to a kind  of  angular momentum operator. In fact, we  can introduce a differential realization of $( \hat y^i,\hat f^i)$ as operators acting  on the space of functions $\Psi(f^1,f^2)$ in the form
\be
\hat y^i  \Psi = \frac 1\kappa \bigl( (  f^1)^2+  (  f^2)^2\bigr) \frac{\partial \Psi}{\partial f^i}\, ,\qquad \hat f^i  \Psi = f^i\Psi ,
\label{ck}
\ee
which can be interpreted as an action on a classical 2D space with momentum-type coordinates $(f^1,f^2)$. Finally, the action of the remaining quantum coordinate $\hat y^-$ on $\Psi$ is straightforward by considering  the Casimir operator $\hat C$ (\ref{ci}):
\be
\hat y^-  \Psi = \hat C \bigl( (\hat f^1)^2+  (\hat f^2)^2 \bigr)\Psi= C  \left( ( f^1)^2+  (  f^2)^2 \right)\Psi ,
\label{cl}
\ee
where $C$ is the eigenvalue of the operator $\hat C$ acting on $\Psi$.

Surprisingly enough, the representation (\ref{ck})  suggests    the following definition  for  two pairs of quantum canonical variables  $(\hat q^i,\hat p^i)$ in terms of the quantum worldline coordinates $( \hat y^i,\hat f^i)$:
\be
\hat q^i := \left ( (\hat f^1)^2+  (\hat f^2)^2\right )^{-1}\,  \hat y^i ,\qquad \hat p^i:=\hat f^i,\qquad    \bigl[ \hat q^i ,  \hat p^j\bigr]=\frac{1}{\kappa}\,\delta_{ij} .
\label{cm}
\ee
 Consequently, the representation  (\ref{ck}) and the definition (\ref{cm}) lead to the  usual realization 
 \be
\hat q^i  \Psi = \frac 1\kappa \, \frac{\partial \Psi}{\partial f^i}\, ,\qquad \hat p^i  \Psi = f^i\Psi ,
\label{cn}
\ee
on functions $\Psi(f^1,f^2)$ defined on a classical momentum space, where now the quantum deformation parameter $\kappa$ plays the same role as the Planck constant $\hbar$ in ordinary quantum mechanics. Moreover, for the fifth quantum worldline coordinate $\hat y^-$ we have that
\be
\bigl[ \hat y^- ,  \hat p^i\bigr]=0\, ,
\qquad
\bigl[ \hat y^- ,  \hat q^i\bigr]= - \frac{2\,C}{\kappa}\, \hat p^i\, ,
\qquad i=1,2 \, ,
\ee
for a given realization characterized by $C$.
Therefore, we can say that the noncommutative space of light-like worldlines~\eqref{ch} can be mapped to a non-central extension, generated by $\hat y^-$, of the direct sum of two Heisenberg--Weyl algebras $(\hat q^i,\hat p^i)$. (As a side remark, note that light-ray noncommutativity arises in spinning light propagation \cite{DHH2006spinning}.) 

At this point we recall that the 6D  noncommutative space of time-like lines constructed in~\cite{BGH2019worldlinesplb} was shown to be generated by the direct sum of three Heisenberg--Weyl algebras, and this enabled the analysis presented in~\cite{BGGM2021fuzzy} of the fuzziness properties of events defined as the crossing of light-like quantum geodesics. In the light-like case we are dealing with a 5D quantum space, whose fuzziness could be studied in a similar manner by taking into account the mapping~\eqref{cm}. We plan to face this analysis in a forthcoming paper. 
 
For the sake of completeness, a few words on the noncommutative Minkowski spacetime arising from the light-like $\kappa$-deformation and its relation with the previous construction seem to be appropriate. This space can be constructed in terms of the Minkowskian null-plane coordinates by considering the Poincar\'e group element obtained from the $5\times 5$  matrix realization (\ref{af}) in the form
 \be
G_\mathcal{M}= \exp\left( {x^+ \rho(P_+)} \right) \exp\left({x^1 \rho(P_1)}  \right)\exp\left( {x^2 \rho(P_2)}  \right)  \exp\left( {x^- \rho(P_-)} \right) H_\st  ,
 \label{cb}
 \ee
where
\be
 H_\st=\exp\left({e'^1 \rho(E_1)}  \right)\exp\left( {e'^2 \rho(E_2)} \right)   \exp\left({f'^1 \rho(F_1)}  \right)\exp\left( {f'^2 \rho(F_2)} \right)     \exp\left({\xi'^3 \rho(K_3)} \right)  \exp\left({\phi'^3 \rho(L_3)} \right)  
\label{cc}
\ee
 is the Lorentz subgroup ${\rm SO}(3,1)$ that plays the role of  the isotropy subgroup  of the point $O=(1,0,0,0,0)$,  which is taken as the origin of the (3+1)D  Minkowski spacetime $\mathcal{M}  =G/H_\st$.

From (\ref{cb}), the left- and right-invariant vector fields on $G$ are obtained, and the Sklyanin  bracket (\ref{bb}) can be computed in terms of the $r$-matrix $r_\nullplane$~\eqref{rlight}. Afterwards, the Poisson structure $\pi$ among the four null-plane Minkowski coordinates $x= (  x^\pm,  x^i ) $ is obtained as the canonical projection of the Sklyanin bracket onto these coordinates. The resulting Poisson structure is linear, and its quantization is straightforward leading to the noncommutative null-plane Minkowski spacetime given by 
\be
[\hat x^{i},\hat x^{+}] =  \frac {2}{\kappa}\,  \hat x^{i} , \qquad    [\hat x^{-},\hat x^{+}] =  \frac {2}{\kappa}\,  \hat x^{-} ,\quad\  i=1,2.
\label{ncslight}
\ee

Notice that  the classical spacetime coordinates $x=(     x^\pm,  x^i )$  in $G_\mathcal{M}$ (\ref{cb}) can be expressed in terms of  those in $G_\mathcal{L}$ (\ref{ce})  as
\bea
&& x^+=y^+,\qquad x^i=y^i +y^+ f^i , \nonumber\\[2pt]
&&x^-= y^- +y^1 f^1+ y^2 f^2 + \tfrac 12 \,y^+\!\left( (f^1)^2+  (f^2)^2   \right) .
\label{cg}
\eea
Obviously, the linearization of~\eqref{cg} gives $ (   x^\pm,  x^i )\equiv  ( y^\pm,  y^i )$.
 Note  also  that if we compute the Poisson brackets for $(x^\pm, x^i)$ by starting from (\ref{chpois})  we recover the linear Poisson structure for the Poisson spacetime (\ref{ncslight}), as it should be. Therefore, the nonlinear map~\eqref{cg} will also provide the starting point for the study of the link between the noncommutative spaces of points~\eqref{ncslight} and light-like geodesics~\eqref{ch} arising from the light-like $\kappa$-deformation.


\section{Concluding remarks}

The definition and explicit construction of the quantum analogue of geodesics on a noncommutative space is indeed an interesting problem from both the mathematical and physical viewpoints. In particular, the case of null geodesics on a noncommutative Minkowski spacetime provides a first toy model that will be (hopefully) useful in order to describe some expected features of the propagation of massless particles on quantum Minkowski spacetimes. 

In this paper we have approached this problem by  constructing explicitly the 5D  noncommutative space of null geodesics arising from the light-like $\kappa$-deformation of the (3+1) Poincar\'e group through the following procedure: firstly, the space of geodesics is constructed as a  homogeneous space  $\mathcal{L}$ of the Poincar\'e group, secondly this space is endowed with a Poisson-noncommutative structure which is provided by the $r$-matrix defining the chosen deformation, and finally the quantization of this Poisson structure is performed, thus obtaining the noncommutative space. We emphasize that, within this framework, each geodesic is represented as a point in the homogeneous space  $\mathcal{L}$, and a similar thing happens within the noncommutative setting here presented. 

We have focused on the particular class of Hopf algebra deformations of the Poincar\'e group given by the so-called $\kappa$-deformations, which can be naturally splitted into time-, space- and light-like cases. Remarkably enough, we have found that only the light-like (or null-plane) $\kappa$-deformation can be used to provide a noncommutative space of light-like worldlines. This could seem to be a natural fact, but it is important to emphasize that these three quantum deformations are different Hopf algebras, and this result points in the direction of considering the complete family of $\kappa$-deformations of the Poincar\'e algebra as the proper Planck-scale symmetry, which has to be specialized to particles with different masses, in the same manner that different particles carry different irreducible representations of the Poincar\'e group. To the best of our knowledge,  this viewpoint has not been considered in the literature yet, and its consequences should be elaborated in more detail.

Also, this mathematical compatibility issue between a chosen quantum deformation and a specific homogeneous space in order to construct the noncommutative version of the latter is worth to be analysed. In particular, the coisotropy condition~\eqref{coisotropy} of the chosen Lie bialgebra with respect to the isotropy subalgebra of the classical homogeneous space is the keystone for this analysis. Specifically, we can  consider the three $r$-matrices given in Table~\ref{table1}, since each of them leads to a Lie bialgebra structure $(\mf g,\delta)$ with commutator given by (\ref{ba}) which, in turn, allows us to analyse whether the coisotropy relation (\ref{coisotropy})  is  fulfilled with respect to the four isotropy subalgebras (\ref{ai}). 

It is straightforward to prove by direct  computation that the three classical $r$-matrices defining the $\kappa$-deformations do satisfy the coisotropy condition for the  construction of a Poisson homogeneous Minkowski spacetime, \emph{i.e.} $\delta(\mf h_{\st})\subset \mf h_{\st}\wedge \mf g$ in all the cases. This means that three noncommutative Minkowski spacetimes can be explicitly constructed from the $\kappa$-Poincar\'e deformations. In fact, all of them can be found in the previous literature, and their non-vanishing commutation relations in terms of the corresponding quantum spacetime coordinates $\hat x^\mu$ (dual to $P_\mu$) are given by the linear relations (see~\cite{Maslanka1993,Rnullplane97,GH2021symmetry}, respectively)
 \be
\begin{array}{lll}
r_\timek\!: & \quad    [\hat x^{a},\hat x^{0}] = \frac {1}{\kappa}\, \hat x^{a}   ,&\quad a=1,2,3. \\[4pt]
r_\spa\!: &\quad [\hat x^{0},\hat x^{3}] =  \frac {1}{\kappa}\,  \hat x^{0} , &\quad
[\hat x^{1},\hat x^{3}] =  \frac {1}{\kappa}\,  \hat x^{1} , \qquad [\hat x^{2},\hat x^{3}] =  \frac {1}{\kappa}\,  \hat x^{2}   .\\[4pt]
r_\nullplane\!: &  \quad [\hat x^{i},\hat x^{+}] =  \frac {2}{\kappa}\,  \hat x^{i} , &\quad    [\hat x^{-},\hat x^{+}] =  \frac {2}{\kappa}\,  \hat x^{-} ,\quad\  i=1,2.
\end{array}
\label{bd}
\ee
Therefore, they are noncommutative spacetimes of Lie-algebraic type and, moreover, all of them are isomorphic as Lie algebras.
  
 Likewise, Table~\ref{table2} shows that the three $r$-matrices  enable the construction of noncommutative spaces of time- and space-like geodesics associated with isotropy subalgebras  $\mf h_\til$  and $\mf h_\sil$ (\ref{ai}), respectively.  However, observe that the coisotropy condition  (\ref{coisotropy})  is trivial $(\delta(\mf h)=0)$  for $r_\timek$ with $\mf h_\til$  and for $r_\spa$ with $\mf h_\sil$, and 
hence  the   Poisson-subgroup condition  (\ref{coisotropy2}) is satisfied in the cases when the type of deformation and the type of homogenous space fit. We recall that the time-like case has been studied at the Lie bialgebra level for   $r_\spa$ and  $r_\timek$    in \cite{GH2021symmetry} (c.f.~classes A and C, respectively), and the structural differences between the two conditions  $\delta(\mf h_\til)\subset \mf  h_\til\wedge \mf g$ and $\delta(\mf h_\til)=0$ have been shown. Finally, as we have discussed in section~\ref{s4}, the construction of noncommutative spaces of light-likes lines is precluded for both the time-like and space-like deformations. Therefore, the only quantum $\kappa$-deformation which  can provide  a noncommutative counterpart of the four homogeneous spaces of points and lines is the light-like $\kappa$-Poincar\'e algebra.

To end with, we comment on two open problems that are worth to be worked out in the near future. Firstly, we recall that the complete classification of  those (3+1)D quantum Poincar\'e and (A)dS algebras that  present a quantum Lorentz subgroup has been recently obtained in~\cite{BGH2021lorentz},
where it has been shown that all of them  
fulfill the  coisotropy condition (\ref{coisotropy}) for the Minkowski spacetime (thus with respect to $ \mathfrak h_\st$). We stress that $\kappa$-deformations are not included within this class of deformations and therefore the analysis of the coisotropy condition for the three spaces of geodesics (\ref{ai}) that are covariant under these novel quantum Poincar\'e groups deserves a further study. Secondly, we recall that the  time-like $\kappa$-deformation of the (A)dS algebra (with nonvanishing cosmological constant $\Lambda$) was explicitly constructed in~\cite{BHMN2017kappa3+1}, and its associated noncommutative spacetime has also been  presented in~\cite{BGH2019kappaAdS3+1}. Therefore, the methodology here proposed can also be  applied to the case of the homogeneous spaces of worldlines associated to this and to the previously mentioned quantum deformations of the (A)dS groups, thus providing a novel approach to the interplay between the cosmological constant and the Planck scale deformation parameter~\cite{BGMsymmetry} based on the properties of quantum worldlines.

  
 \begin{table}[t]
{\small
\caption{\small    Coisotropy condition (\ref{coisotropy}) for the three classical $r$-matrices in Table~\ref{table1} with respect to the three different  isotropy subalgebras (\ref{ai}) that define the time-like (tl), space-like (sl) and light-like (ll) homogeneous spaces of worldlines.}
\label{table2}
  \begin{center}
\noindent 
\begin{tabular}{l l l l}
\hline

\hline
\\[-8pt]
\qquad\qquad\  & $\mf h_\til$& $\mf h_\sil$& $\mf h_\lil$ \\[4pt]
\hline
\\[-8pt]
$r_\timek$ & $\delta(\mf h_\til)=0$& $\delta(\mf h_\sil)\subset \mf h_\sil\wedge \mf g\quad$ & $\delta(\mf h_\lil)\not\subset \mf h_\lil\wedge \mf g$\\[4pt]
$r_\spa$ &  $\delta(\mf h_\til)\subset \mf h_\til\wedge \mf g\quad$  &  $\delta(\mf h_\sil)=0$& $\delta(\mf h_\lil)\not\subset \mf h_\lil\wedge \mf g$  \\[4pt]
$r_\nullplane$ & $\delta(\mf h_\til)\subset \mf h_\til\wedge \mf g$& $\delta(\mf h_\sil)\subset \mf h_\sil\wedge \mf g$& $\delta(\mf h_\lil)\subset \mf h_\lil\wedge \mf g$ \\[4pt]
\hline

\hline
\end{tabular}
 \end{center}
}
 \end{table} 
 

\section*{Acknowledgements}

This work has been partially supported by   Agencia Estatal de Investigaci\'on (Spain)  under grant  PID2019-106802GB-I00/AEI/10.13039/501100011033. 
 The authors would like to acknowledge the contribution of the   European Cooperation in Science and Technology COST Action CA18108.


\small


\end{document}